\documentclass[sigconf,nonacm]{acmart}
\AtBeginDocument{%
  }

\setcopyright{acmlicensed}
\copyrightyear{2026}
\acmYear{2026}
\acmDOI{XXXXXXX.XXXXXXX}

\usepackage{tikz}
\usetikzlibrary{shapes.geometric, arrows.meta, positioning, fit, backgrounds, calc}
\usepackage{booktabs}
\usepackage{framed}
\newcounter{prompt}
\newenvironment{promptbox}[1]{%
  \refstepcounter{prompt}%
  \begin{framed}\noindent\textbf{Prompt~\theprompt: #1}\par\medskip
  \footnotesize\ttfamily\raggedright
  \setlength{\parindent}{0pt}\setlength{\parskip}{0.1em}}
  {\end{framed}}
\usepackage{enumitem}

\begin{document}

\title{Beyond Smart Homes: Characterizing IoT Vulnerabilities Across Sectors, Vendors, and Products}

\author{Tina Rezaei}
\email{t.rezaei@utwente.nl}
\affiliation{%
  \institution{University of Twente}
  \city{Enschede}
  \country{The Netherlands}
}

\author{Suzan Bayhan}
\email{s.bayhan@utwente.nl}
\affiliation{%
  \institution{University of Twente}
  \city{Enschede}
  \country{The Netherlands}
}

\author{Andrea Continella}
\email{a.continella@utwente.nl}
\affiliation{%
  \institution{University of Twente}
  \city{Enschede}
  \country{The Netherlands}
}

\author{Jeroen van der Ham-de Vos}
\email{j.vanderham@utwente.nl}
\affiliation{%
  \institution{University of Twente}
  \city{Enschede}
  \country{The Netherlands}
}

\author{Roland van Rijswijk-Deij}
\email{r.m.vanrijswijk@utwente.nl}
\affiliation{%
  \institution{University of Twente}
  \city{Enschede}
  \country{The netherlands}
}

\renewcommand{\shortauthors}{Rezaei et al.}

\begin{abstract}
 The Internet of Things (IoT) has expanded beyond consumer smart homes to sectors, such as industrial systems, healthcare, and smart cities, broadening the IoT attack surface. Prior empirical work has focused on studying aggregate IoT vulnerability characteristics or specific protocols, so patching is directed by aggregate security guidance. However, if dominant weaknesses differ substantially across sectors or products, such uniform guidance misdirects the remediation effort. To address this gap, we present a fine-grained, data-driven analysis of IoT vulnerabilities using a dataset of more than 15,000 IoT-specific CVEs. We design a pipeline to enrich noisy CPE (Common Platform Enumeration) metadata, normalizing vendors and products, then assigning product type and sector labels using lexical matching, fuzzy clustering, and LLM-assisted classification. We find meaningful differences in the weakness profile of IoT sectors. For example, healthcare is dominated by authentication and credential weaknesses, while automotive faces memory issues. More importantly, we find that weakness profiles are more similar across different product types from the same vendor than across the same product type built by different vendors. This suggests vendor-level factors such as shared code and development practices dominate weakness inheritance, making a vendor's disclosure history informative about its other products. Our findings suggest weakness classes should be used to prioritize remediation efforts.
\end{abstract}

\begin{CCSXML}
<ccs2012>
    <concept>
    <concept_id>10002978.10003006.10011634</concept_id>
    <concept_desc>Security and privacy~Vulnerability management</concept_desc>
    <concept_significance>500</concept_significance>
    </concept>

    <concept>
    <concept_id>10010147.10010257.10010258.10010260.10003697</concept_id>
    <concept_desc>Computing methodologies~Cluster analysis</concept_desc>
    <concept_significance>500</concept_significance>
    </concept>

     <concept>
    <concept_id>10002978.10003001.10003003</concept_id>
    <concept_desc>Security and privacy~Embedded systems security</concept_desc>
    <concept_significance>100</concept_significance>
    </concept>    
</ccs2012>

\end{CCSXML}

\ccsdesc[500]{Security and privacy~Vulnerability management}
\ccsdesc[500]{Computing methodologies~Cluster analysis}
\ccsdesc[100]{Security and privacy~Embedded systems security}

\keywords{IoT security, CVE, Sector analysis, LLM-assisted classification, Weakness inheritance}

\maketitle

\section{Introduction}

The Internet of Things (IoT) now spans nearly every sector of the modern economy \cite{xenofontos2021consumer}. Smart devices enable remote healthcare monitoring for patients, drive infotainment functions in vehicles, and instrument traffic and surveillance in cities. Each new deployment expands the attack surface, and a vulnerability in a widely reused component can cascade across vendors, products, and sectors \cite{zhao2023one}. Therefore, understanding the vulnerability landscape of IoT systems has become a critical research priority for designing effective and targeted mitigation strategies.

Prior empirical studies have concentrated either on aggregate trends and broadly characterizing the IoT threat landscape \cite{mouratidis2018security}, or on narrow attack vectors and individual device categories \cite{tagliaro2024large, costin2014large, zhao2020large}, mostly in consumer smart home settings \cite{alrawi2019sok}. Despite a large body of work on IoT security \cite{swamy2020empirical,neshenko2019demystifying}, the IoT vulnerability landscape remains poorly characterized at finer-grained levels, such as sector and product type levels. For example, a hospital containing infusion pumps and patient monitors risks unauthorized access to sensitive data, while automotive chipsets that are built on embedded C/C++ risk memory-safety issues. Prior work cannot reveal how the threat surface differs in various sectors, whether vendors building similar products are exposed to similar weaknesses, and how widely individual vulnerabilities propagate across the supply chain.

Understanding these aspects at scale is nontrivial due to incomplete and inconsistent vulnerability data. The National Vulnerability Database~(NVD)  is the most comprehensive public source of disclosed vulnerabilities, and its Common Platform Enumeration~(CPE) metadata links each CVE to affected vendors and products. However, the current state of this knowledge is ambiguous, semantically inconsistent, and incomplete \cite{jiang2025vulcpe}: the same vendor appears under many name variants, and CPE entries carry no notion of product category or deployment sector, so a string such as ``sm7315-firmware'' (a Qualcomm chipset) cannot be mapped to a meaningful product type or sector. Manual labeling is infeasible across tens of thousands of CVEs, and traditional text clustering struggles because CPE identifiers are short and carry little natural-language signal. However, recent LLM-based clustering methods overcome this limitation by enabling semantic-aware clustering.

In this paper, we present a first systematic fine-grained analysis of IoT vulnerabilities across sectors, vendors, and products. We use a recent dataset of over 15,000 IoT CVEs~\cite{rezaei2025lliot}, which is more accurately scoped to IoT-specific vulnerabilities than prior datasets~\cite{janiszewski2021automatic, chen2024empirical}. We develop a semi-supervised pipeline that combines lexical normalization, fuzzy matching, and two-step LLM-based clustering to assign every affected product a product type label and a sector. We produce consistent, human-interpretable labels at scale. We verify the correctness of the generated labels by manual inspection.

Then, we use this enriched dataset to study three research questions: (RQ1) How are IoT vulnerabilities distributed across sectors and how do the threat surfaces of sectors differ from one another? (RQ2) How do IoT weakness profiles vary within a vendor's products and across vendors building the same kind of product? and (RQ3) To what extent do vulnerabilities propagate across the supply chain in terms of shared vendors and products?

\section{Related work}
Here, we review prior studies that characterized IoT vulnerabilities and classified IoT devices, as well as LLM-based text clustering approaches, since our pipeline needs to cluster opaque CPE strings.

\noindent\textbf{IoT vulnerability analysis \& product classification}. Falco et al.~\cite{falco2018iiot} evaluate SCADA software risk using CVE and CWE data drawn from ICS-CERT and NVD. Several studies have proposed classifications of IoT devices (products) from different perspectives.
Bugeja et al. \cite{bugeja2018functional} presented a
classification of smart home devices, categorizing them into eight clusters based on their functionality (e.g., energy management, entertainment, health and wellness, household). In follow-up work, Bugeja et al.~\cite{bugeja2020} used consumer IoT device specifications and an unsupervised machine learning model to cluster IoT devices based on their data-collection capabilities into four groups: app-based accessors, watchers, location harvesters, and listeners. Kumar et al. \cite{kumar2019all} categorized IoT devices into eleven classes depending on their type (e.g., wearable, surveillance, home voice assistant). IoTSentinel~\cite{miettinen2017iot} and HomeSnitch~\cite{oconnor2019homesnitch} classify smart home devices from their traffic fingerprints. The proposed classifications do not cover the whole IoT spectrum throughout all sectors and mostly focus on one layer of the IoT ecosystem, e.g., device, lacking other layers such as communication and application.

\noindent\textbf{Text clustering}. Recent studies have proposed various LLM-based clustering algorithms, outperforming traditional clustering algorithms' performance by enabling semantic-aware clustering. Feng et al.~\cite{feng2024llmedgerefine} propose LLMEdgeRefine, which initializes clusters with K-means and refines boundary-ambiguous points via LLM reassignment. Their approach requires a predefined number of clusters and fine-tuned embeddings, which do not fit our setting where the number of CVE product categories is unknown. Pattnaik et al.~\cite{pattnaik2024improving} propose a hierarchical clustering that combines multi-view LLM-guided representations. Their approach requires rich document-level text for bottom-up agglomeration, which short and opaque CPE identifiers do not provide. Huang and He~\cite{huang2025text} reformulate text clustering as a classification task without needing to know the number of clusters in advance. They first generate candidate labels in mini-batches and merge generated labels, and then assign samples to merged labels. We adapt this approach to handle our scale and label-granularity requirements, as described in Section~\ref{sec:methodology}.

\section{Methodology}
\label{sec:methodology}
We perform a semi-supervised classification of IoT products present in IoT CVEs, assigning each a deployed sector and a product type. For sector classification, we use the IoT market segmentation reported by Statista\footnote{https://www.statista.com/forecasts/1444694/revenue-internet-of-things-market-for-different-segments-worldwide/}, which defines seven sectors: Consumer IoT, Industrial IoT, Smart cities, Smart finance, Healthcare IoT, Automotive IoT, and others. This set represents the market-facing IoT deployments and covers the major IoT environments.

In contrast, providing a finite list for product type classification is not possible. The diversity of products across sectors and vendors is too large, and available lists in the literature are incomplete. Therefore, we take a data-driven approach. We first pre-process CPE data to merge vendor name variants and normalize product names in the whole dataset using lexical normalization and fuzzy clustering. Then we use a two-step LLM-based approach for product type label generation and aggregation. Fig. \ref{fig:methodology} in Appendix \ref{app:methodology} gives an overview of the complete pipeline. In the rest of this section, we describe the adopted dataset and explain our methodology in detail.

\noindent\textbf{Dataset}. As our source of IoT vulnerabilities, we use the LLIoT dataset \cite{rezaei2025lliot}, containing 15,116 IoT CVEs along with their CPE metadata and Common Weakness Enumeration (CWE) assignments. The dataset is built by classifying all CVEs from 2013 to 2024 using an LLM-assisted pipeline guided by a four-layer IoT definition (device, communication, application, and cloud). The dataset covers IoT CVEs more comprehensively than previous datasets \cite{chen2024empirical, janiszewski2021automatic}. Since only 14,696 of them have both vendor and product metadata, our analysis focuses on this subset.  Note that each CVE might affect various vendors and products and each might be deployed in one or more sectors. Each CPE record specifies an affected vendor, product, and a \texttt{part} field indicating the layer at which the product operates: \texttt{h} (hardware), \texttt{o} (operating system), or \texttt{a} (application).   

Vulnerabilities affecting the same product type can differ substantially depending on the attacked layer. For instance, a camera presents a different attack surface in its firmware (\texttt{o}) than in its web management application (\texttt{a}), and mixing the two would obscure meaningful differences in their weakness profiles. Appendix. \ref{app:sample} shows a sample NVD record with related CVE description, CWE, and CPE records.

\subsection{Data preprocessing}
Raw CPE metadata is inconsistent and redundant. The same vendor or product may appear under multiple name variants due to abbreviations, misspellings, or versioning suffixes. Without normalization, these variants inflate entity counts and decrease clustering quality. Therefore, first we apply a multi-step normalization to vendors and products, which also reduces downstream LLM cost.

\noindent\textbf{Vendor normalization}. First, we normalize vendors using a regular expression (lower casing, removing punctuation, and corporate suffixes). We also apply fuzzy string matching using token-based similarity to merge variants that differ only in minor variations, e.g., 'philips' and 'phillips'. 

Some vendors cannot be merged due to abbreviations or large textual differences between a company's short and full names (e.g., \textit{BD} vs. \textit{Becton Dickinson}). Such cases cannot be resolved with traditional lexical methods. We address these with an LLM-based deduplication pass. We first sort vendors alphabetically and submit them to the LLM in per-letter batches, asking the LLM to merge the same vendors. Name variants of the same vendor almost always share the same initial character. This approach keeps batches small enough to avoid context length issues while placing the most likely variants together.

\noindent\textbf{Product normalization}. We normalize products per vendor instead of globally. Global product normalization risks incorrectly merging similarly named products from different vendors that refer to distinct physical devices. We apply regular expression cleaning to strip generic suffixes (e.g., \textit{standard}, \textit{module}) and punctuation.

\subsection{LLM-assisted classification}

Following LLM-based clustering approaches in the literature, we utilize LLMs to generate semantic-aware, interpretable product type labels. We divide our data into mini-batches, generate candidate labels per batch, and merge them to reduce redundancy. We also provide the LLM with previously generated labels during batch processing to encourage label reuse and consistency across batches, which prevents fragmentation of semantically similar labels. 

\noindent\textbf{Label generation}. 

The same (vendor, product) pair typically appears across many CVEs. Therefore, we collapse this redundancy by selecting unique (vendor, product) pairs before LLM processing, keeping the longest CVE description among all CVEs associated with each pair to provide the richest available context for tagging. The resulting label is then propagated back to every original (CVE, product) row in the output, ensuring every CVE record receives a label while reducing LLM calls. 

Instead of inputting the entire data into LLMs, which is impractical due to context length limitations, we divide the resulting pairs into mini-batches for LLM processing. Rather than dividing the pairs into fixed-size batches naively, we divide them hierarchically, based on vendor size and product type diversity. First, we group pairs by vendor to preserve labeling consistency within each vendor. Large vendors, those with at least $m$ distinct affected products, are processed individually, while small vendors are pooled together until the pool reaches m distinct products, ensuring sufficient product diversity for meaningful clustering. Second, within each vendor or vendor pool, we construct batches of at most $m$ distinct products per LLM call, further preventing the context limit problem. 

During batch processing, the LLM is provided with the list of product type labels generated in previous batches for the same vendor or pool, allowing it to reuse existing labels rather than generate redundant variants of the same product type. We also seed each vendor's first batch with a small set of manually crafted common IoT product type labels (e.g., \textit{camera}, \textit{router}, \textit{PLC}, \textit{SCADA}) to better guide the LLM for the intended granularity level of the labels. Then we prompt the LLM to assign a product type and a sector label for each (vendor, product) pair. For reference, the prompt we used for this analysis is included in the appendix (Appendix. \ref{app:prompts}).

\noindent\textbf{Merging labels}. Once all candidate product type and sector labels are generated across batches, we aggregate and merge product type labels to remove duplicates and synonyms. We perform this merging step per sector rather than globally. The previous step determines the affected sector for each vendor and product pair. Therefore, for generated product type labels within a sector, we sort them alphabetically and split them into batches of up to 30. Then we prompt LLMs to merge synonyms. Merging is not independent across batches. Each batch receives the set of merged labels already generated in all previous batches in the same sector, and the prompt instructs the LLM to reuse an existing label exactly when the input fits, creating a new merged label only when none of the existing labels is a reasonable match. This provided cross-batch context prevents fragmentation where equivalent labels (e.g., IP Camera and Network Camera) appear in separate batches. The prompt we used for this analysis is included in the appendix (Appendix. \ref{app:prompts}).

\section{Results}

We used gpt-5o-nano as the LLM for label generation and aggregation. We set the batch size $m$ during label generation to 5. This number is selected practically based on the prompt size to avoid context length limit issues. Evaluating the effect of batch size on the quality of the generated labels is beyond the scope of this research. The label generation and merging labels processes incurred a cost of approximately USD\$ 10 together.

\subsection{Normalization effectiveness}
Regarding vendor normalization, lexical normalization and fuzzy matching reduced the number of unique vendors from 2,097 to 2,082. This number was further reduced to 1,984 after the LLM deduplication pass. Per-vendor product normalization reduced unique products from 26,720 to 10,238. After these normalization steps, the dataset comprises 85,985 distinct (CVE-ID, vendor, product) triplets, collapsing these to unique (vendor, product) pairs results in 10,416 pairs for LLM processing.

\subsection{Classification quality}
We validate the quality of our classification along two dimensions, corresponding to the two LLM-driven decisions in our pipeline:

\noindent\textbf{Sector assignment.} We sample 200 distinct (vendor, product type) pairs, stratified across the seven sectors. We manually verify that the assigned sector matches the product's primary deployment context for each sample. 

\noindent\textbf{Product type labels.} We sample 200 pairs. We divide data into 3 buckets (large, medium, small) based on vendor size and sample 67 from each bucket. For each sample, we manually verify whether the assigned product type label accurately describes the product.

We asked two reviewers (both PhD students with expertise in cybersecurity) to independently evaluate these samples. 

Reviewers determine whether each sample's assigned sector label and product type label are correct. We calculate accuracy as the fraction of samples where both reviewers agreed that the label is correct. The obtained accuracy for product type labels and sector labels is 84\% and 88.6\%, respectively. Modeling each reviewed label as a Bernoulli trial (correct vs.\ incorrect), the 95\% Wilson confidence interval for product type labels ($k = 168$ of $n = 200$) is [78.3\%, 88.4\%], and for sector labels ($k = 177$ of $n = 200$) is [83.3\%, 92.2\%]. This result shows that our LLM pipeline produces largely reliable labels. 

\begin{figure}
    \centering
    \includegraphics[width=0.8\linewidth]{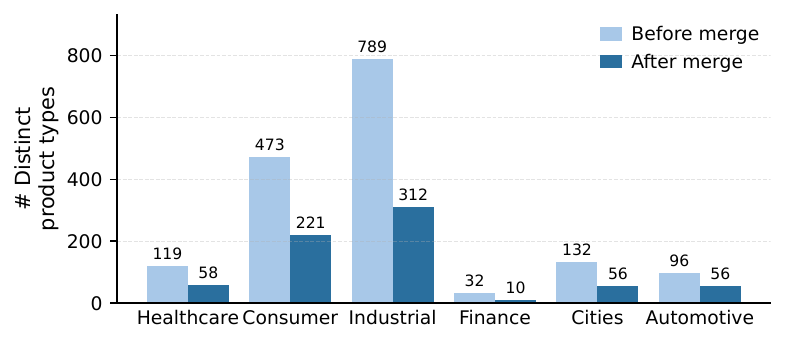}
    \caption{Number of distinct product types in each sector before and after the merge step.}
    \label{fig:profuct_types_before_after}
\end{figure}

Fig.~\ref{fig:profuct_types_before_after} reports the number of distinct product type labels per sector before and after the merging step, consolidating near-duplicate labels, specifically in the Industrial (789 to 312) and Consumer (473 to 221) sectors.

\subsection{Vulnerability distribution across sectors}

To answer RQ1, we first show the temporal pattern of IoT vulnerabilities across sectors. Then we illustrate the most common CWEs and top product types across sectors, cross referencing each sector's CWE profile to that sector's product types.

\begin{figure}
    \centering
    \includegraphics[width=0.85\linewidth]{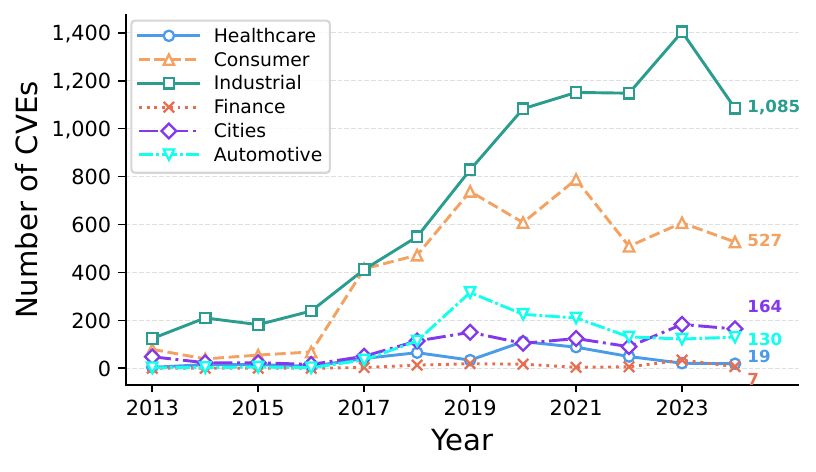}
    \vspace{-10pt}
    \caption{IoT CVEs temporal growth by sector.}
    \label{fig:iot_sector_temporal}
\vspace{-10pt}
\end{figure}

\begin{figure*}
    \centering
    \includegraphics[width=\linewidth]{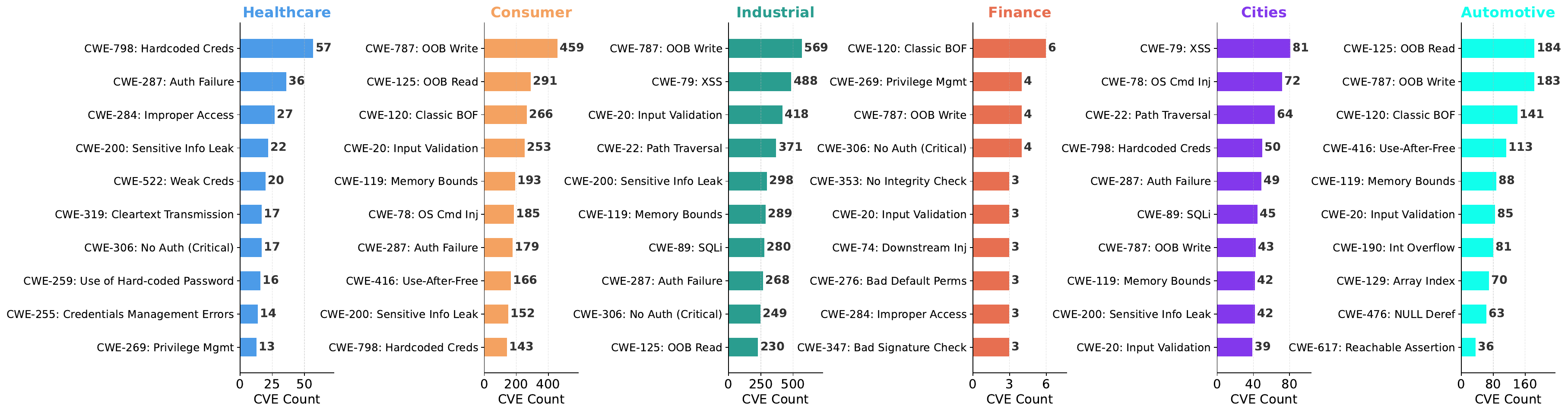}
    \caption{Top 10 CWEs in each sector.}
    \label{fig:sector_cwe}
\end{figure*}

\begin{figure*}
    \centering
    \includegraphics[width=\linewidth]{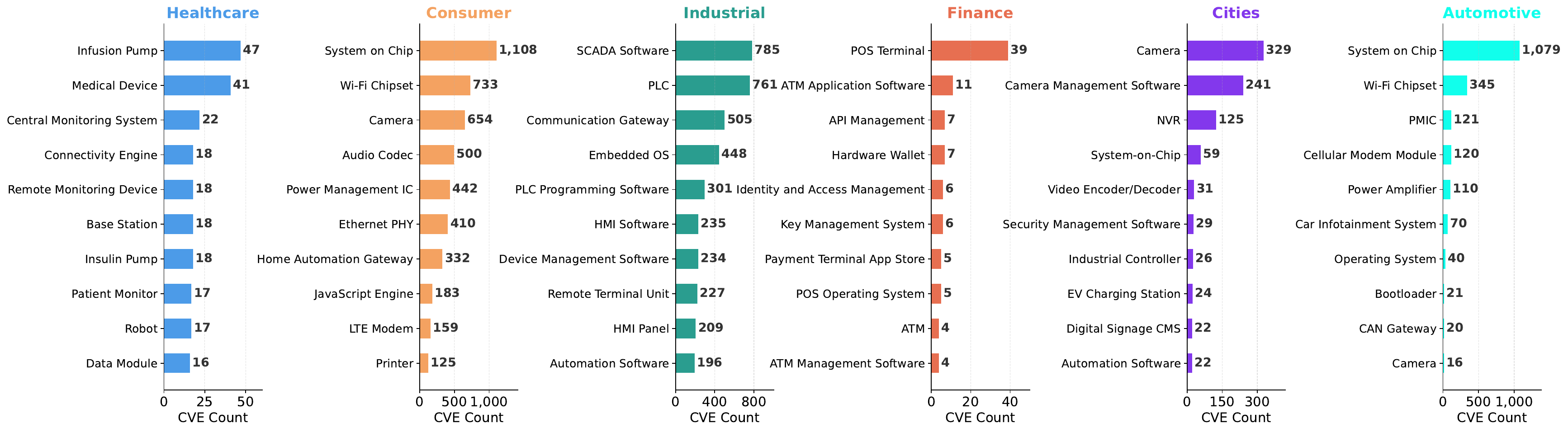}
    \caption{Top products per sector.}
    \label{fig:sector_products}
    \vspace{-10pt}
\end{figure*}

Fig.~\ref{fig:iot_sector_temporal} shows the temporal evolution of IoT vulnerabilities by sector from 2013 to 2024. The paper describing the dataset we have used \cite{rezaei2025lliot} showed that IoT vulnerabilities increased drastically after 2016, probably due to the public release of the Mirai botnet source code and the concurrent expansion of the CVE Numbering Authority (CNA) program. This pattern is also observable for the sectors in Fig.~\ref{fig:iot_sector_temporal}, specifically for the industrial and consumer sectors. Among all sectors, the industrial sector dominates throughout the period, and its growth is pronounced after 2018, reaching 1,085 unique CVEs in 2024 alone, affecting industrial products. Next, the consumer sector is the next most affected, with 527 vulnerabilities in 2024 alone, and yet far above the other remaining sectors. Although the share of vulnerabilities in other sectors in each year is small, we observe an increasing trend in each of them. Among them, automotive IoT and smart cities account for more vulnerabilities.

Next, we investigate the most common weaknesses and top product types in each sector. The most common weaknesses in IoT have been reported to be memory-safety issues, followed by web-related and authentication weaknesses~\cite{rezaei2025lliot}. However, it is not investigated whether this observation holds in each sector. Figs.~\ref{fig:sector_cwe} and \ref{fig:sector_products} show top CWEs and top product types within each sector, revealing markedly different profiles. Healthcare IoT is dominated almost entirely by credential, authentication, and access-control weaknesses. Weaknesses such as hardcoded credentials (CWE-798), authentication failure (CWE-287), improper access control (CWE-284), sensitive info leakage (CWE-200), weak credentials (CWE-522), and sensitive cleartext transmission (CWE-319) collectively account for the majority of its top 10 CWEs.  Memory safety weaknesses do not appear in the healthcare top 10 at all. This profile matches the sector's top product types, including infusion pumps, medical devices, central monitoring systems, patient monitors, and remote monitoring devices. This networked clinical equipment and remote-monitoring products handle highly sensitive personal data. Our observation is consistent with a recent Internet-wide study of exposed healthcare services~\cite{bruggemann2026measuring}, which found that many of them lack authentication. The authors attribute this to protocols that make authentication optional or absent (e.g., DICOM, HL7), services designed for internal networks but being exposed due to incorrect network segmentation, and the risk that security controls (e.g., expired TLS certificates) disrupt communication between medical devices. 

The consumer IoT sector shows a profile dominated by memory safety, with input validation and command injection as secondary contributors. memory-safety issues such as out-of-bounds write (CWE-787), out-of-bounds read (CWE-125), classic buffer overflow (CWE-120), and memory bounds (CWE-119) account for half of its top 10 weaknesses, alongside improper input validation (CWE-20), OS command injection (CWE-78), and authentication failure making up the second half. This pattern reflects the sector's product mix, dominated by system-on-chip (1,108 CVEs), followed by Wi-Fi chipsets, cameras, audio codecs, power management ICs, and other resource-constrained embedded components mostly written in the memory-unsafe C/C++ languages, which causes memory corruption and unvalidated input parsing attack surfaces.  

The industrial sector is similar to the consumer sector in its memory safety dominance,  with out-of-bounds (OOB) write (CWE-787) at the top and memory bounds (CWE-119) and OOB read (CWE-125) also in the top 10. However, web weaknesses follow closely in prevalence: XSS (CWE-79) ranks second, path traversal (CWE-22) fourth, and SQL injection (CWE-89) seventh. We also observe authentication issues in the top 10 weaknesses (CWE-287 and CWE-306). This weakness profile shows heterogeneity of products in this sector, including SCADA software, PLCs, communication gateways, embedded operating systems, and HMI panels and software. Many of these products use low-level memory-unsafe languages and expose web-based management interfaces.

The finance sector, with its substantially smaller CVE count  (its top CWE accounts for only 6 entries), shows a more diffuse distribution across access control, integrity weaknesses, and a few memory and injection issues. However, the low absolute counts make sector-level conclusions less robust here.

Smart cities are dominated by web vulnerabilities, XSS (CWE-79) is first, OS command injection (CWE-78) second, path traversal (CWE-22) third, and SQL injection (CWE-89) sixth. Then, credential and authentication issues (CWE-798) dominate, followed by memory safety weaknesses at the lower end.  This is consistent with the sector's product types, including network cameras, camera management software, NVRs, digital signage CMS, EV charging stations, and security management software, which span both web-facing management consoles and embedded firmware.

The automotive sector presents the most concentrated profile. Its top 10 is almost entirely memory safety, led by out-of-bounds read (CWE-125) and write (CWE-787), followed by classic buffer overflow (CWE-120), use-after-free (CWE-416), and memory bounds (CWE-119). Integer overflow (CWE-190) and improper array index validation (CWE-129) also appear, suggesting systematic issues in numeric handling. This is consistent with automotive reliance on safety-critical, performance-optimized C/C++ code used in system-on-chip (1,079 CVEs), Wi-Fi chipsets, cellular modem modules, and car infotainment systems, causing memory-safety issues.

\subsection{Weakness inheritance}
In order to answer RQ2 and analyze whether weaknesses are inherited across vendors or across product types, we construct a CWE profile for each distinct (vendor, product type, CPE-part) triplet. For each triplet, we collect all CVEs associated with that combination, count how often each CWE class appears, and normalize the resulting vector to sum to 1. The CWE profile is therefore a probability distribution over CWE classes for that (vendor, product, part) triplet.
We then compare CWE profiles under three conditions:
\begin{itemize}[leftmargin=*, itemsep=2pt, topsep=3pt, parsep=0pt]
    \item  Same vendor, different product types: pairs drawn from within a single vendor's portfolio (e.g., D-Link router-firmware vs.\ D-Link camera-firmware). This experiment measures whether individual vendors share similar weaknesses across different product types. 
    \item Same product type, different vendors: pairs drawn from the same product type across vendors (e.g., D-Link router-firmware vs.\ TP-Link router-firmware). This experiment measures whether a specific product type shares similar weaknesses across vendors.  
    \item No-shared: unrelated pairs that share neither vendor nor product type. This is a baseline to show the similarity expected by chance. 
\end{itemize}

We calculate pairwise similarity based on cosine similarity between CWE profiles, applying three constraints to control for bias. First, we restrict the analysis to single-vendor CVEs. Otherwise, different vendors affected by the same CVE artificially increase the similarity score for the same-product type similarity. Second, we require at least three distinct product types per vendor for the same-vendor condition, and at least three distinct vendors per product type for the same-product type condition. Third, for the same vendor condition, since vendors differ widely in size, we compute the median of all similarity scores obtained from product pairs within each vendor, preventing large vendors from dominating the result. Then we plot the obtained similarity scores across all vendors.

\begin{figure}
    \centering
\includegraphics[width=0.85\linewidth]{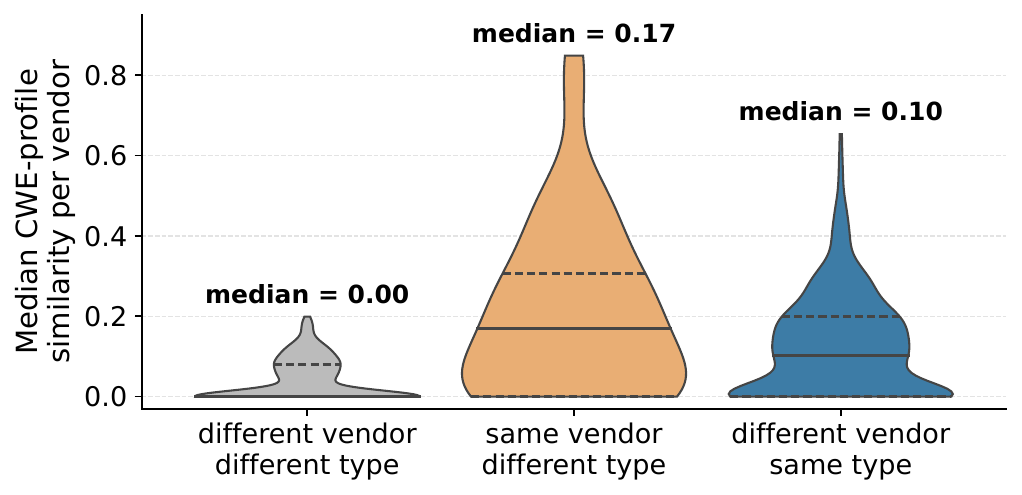}
\vspace{-5pt}

    \caption{CWE profile similarity.}
\label{fig:CWE_profile_similarity}
\vspace{-10pt}
\end{figure}

Fig. \ref{fig:CWE_profile_similarity} shows the CWE profile similarity for the three conditions. The baseline condition shows near-zero similarity across all triplets, suggesting that unrelated products from unrelated vendors have different attack surfaces. In contrast, the same vendor, different product types condition shows substantially higher similarity (median of 0.17), with a wide distribution that exceeds 0.5 for some vendors, indicating that vulnerabilities propagate across a vendor's products despite functional differences. The same product type, different vendors condition shows lower similarity, with a median of 0.10 and in some cases, reaching above 0.5. However, this number still shows a meaningful similarity between products in the same category (e.g., IP cameras from different manufacturers). Although the absolute similarities are modest, they are well above the baseline, and low values are partly expected: CWE profiles are sparse vectors built from a limited number of CVEs, and cosine similarity treats closely related CWEs (e.g., CWE-787 and CWE-119) as unrelated. We compare each condition against the baseline using a one-sided Mann--Whitney U test, and report Cliff's $\delta$ as the effect size (Cliff's 
$\delta$ ranges from $-1$ to $+1$; 0 indicates no tendency for one group to exceed the other, ±1 complete separation). Both elevations are significant with medium effect sizes: same product-type across vendors ($p < 10^{-31}$, Cliff's $\delta = +0.40$) and same vendor across product types ($p = 5 \times 10^{-8}$, $\delta = +0.47$).

\begin{figure}
    \centering
    \includegraphics[width=0.89\linewidth]{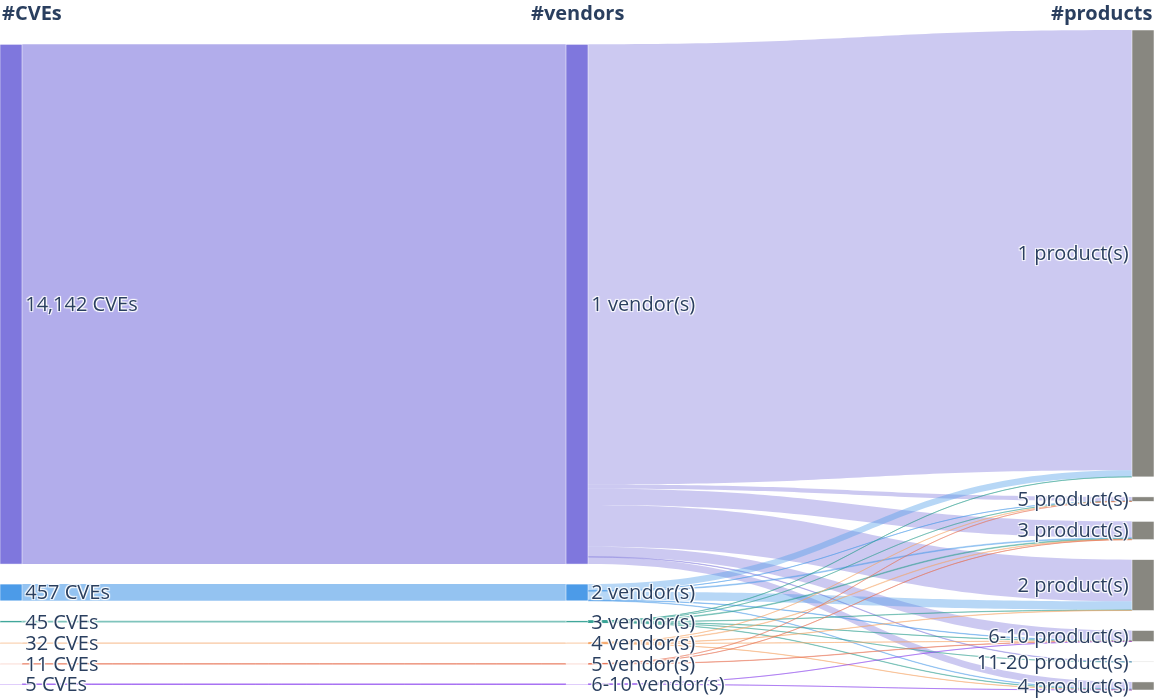}
    \caption{Vulnerability propagation across vendors and products.}
    \label{fig:supply_chain_sankey}
\vspace{-10pt}
\end{figure}

\subsection{Supply chain vulnerability propagation}
To answer RQ3 on how IoT vulnerabilities propagate across the supply chain, we examine how often a single CVE affects multiple vendors and product types. Fig.~\ref{fig:supply_chain_sankey} shows the joint distribution of the number of vendors and number of products affected by each CVE as a Sankey diagram. The left column groups CVEs by the number of vendors they affect, and the right column further splits each group by the number of affected products. Most CVEs affect only a single vendor, and among these, 15\% still propagate across two or more products. For a small fraction of CVEs we observe multiple vendors affected: 457 CVEs affect two vendors, 45 affect three, 32 affect four, 11 affect five, and 5 affect between six and ten vendors. While these multi-vendor and multi-product CVEs represent a small fraction of the total, they propagate simultaneously across several vendors and products, thereby multiplying the effective attack surface and creating widespread exposure.

\section{Discussion and limitations}

Our analysis showed a noticeable difference between threat surfaces in different sectors and also showed that vendors share similar threat surfaces across their products.  Prior aggregate characterizations report that memory safety dominates IoT vulnerabilities, followed by web-related and authentication weaknesses \cite{rezaei2025lliot}. Our sector-level analysis shows this aggregate ranking is largely an artifact of the two largest sectors, consumer and automotive, and holds only partially for industrial. These findings have direct implications for remediation strategies. Rather than applying uniform patching policies across all IoT devices, organizations should prioritize weakness classes aligned with their sector and vendor. For instance, healthcare providers should focus on authentication and credential management vulnerabilities, while automotive manufacturers should prioritize memory-safety issues. Similarly, vendors can use sector-specific and cross-product CWE profiles to guide security investments and set code review priorities. Note that the observed weaknesses might shift over time as patching practices change or new device classes emerge. However, we expect the broad pattern to persist, as it reflects structural and underlying characteristics.

\noindent\textbf{Limitations.} One of our limitations is that our analysis is restricted to only vulnerabilities that were incorporated in the NVD and were assigned CPE metadata and CWE labels. CWE assignment by NVD is itself noisy, and assigned CWEs are uneven across vendors. For example, a vendor whose CVEs were all reported by a single researcher may show an artificially tight CWE profile reflecting the researcher's labeling habits rather than the underlying code. 

Cosine similarity on raw CWE IDs treats similar CWEs (e.g., CWE-787 out-of-bounds write and CWE-119 memory-bounds, which are parent-child in the CWE taxonomy) as completely unrelated. However, we would expect higher observed similarity scores in that case, suggesting our reported numbers are conservative. Despite these limitations, our results demonstrate consistent and meaningful patterns in the attack surface of vendors and sectors.

Classification errors might affect our results, especially for small sectors such as Finance or for rare product types, where adding or removing a few samples can noticeably change the observed CWE profile. For other sectors, the main conclusions are less sensitive to individual labeling errors due to the large number of samples.

\section{Conclusion}
We characterized 14,696 IoT vulnerabilities across sectors, vendors, and product types. Using a semi-supervised LLM-assisted pipeline, we enriched IoT CVEs with sector and product type labels.  Our findings revealed that weakness profiles vary substantially between sectors, suggesting that sector-specific mitigation strategies are necessary. We further revealed that vendors share similar CWE profiles across their products, indicating that vulnerability inheritance is driven by a vendor's codebase and security practices. Although some vulnerabilities propagate across the supply chain through shared components, most remain limited to a single vendor. Our results provide strong signals for organizations and vendors to prioritize remediation efforts based on the weakness classes that they are significantly exposed to.  In future work, we will investigate the root causes of multi-vendor vulnerabilities and how quickly patches propagate through dependency chains.

\bibliographystyle{ACM-Reference-Format}
\bibliography{references}
\appendix

\section{Sample record}
\label{app:sample}
Table~\ref{tab:sample} shows a sample record CVE-2019-15914, which affects several Xiaomi products, listed in its CPE records as firmware (part \texttt{o}) running on the corresponding hardware (part \texttt{h}). The product names (e.g., \texttt{mccgq01lm\_firmware}) are vendor model numbers that give no indication of the device's type or functionality. In fact, these are ZigBee-based consumer IoT devices: a smart plug (ZNCZ03LM), a door/window sensor (MCCGQ01LM), a temperature and humidity sensor (WSDCGQ01LM), and a motion sensor (RTCGQ01LM). The vulnerability stems from how these devices handle the ZigBee Trust Center rejoin procedure, which an attacker can abuse to trigger a denial-of-service (DoS) condition. This example illustrates why lexical matching on CPE strings alone cannot recover product types, motivating our LLM-assisted classification.

\begin{table}[h]
\centering
\footnotesize
\caption{Raw NVD sample record for CVE-2019-15914}
\label{tab:sample}
\begin{tabular}{@{}p{0.2\columnwidth}p{0.74\columnwidth}@{}}
\toprule
\multicolumn{2}{@{}l}{\textbf{Raw NVD record}}\\
\midrule
CVE ID & CVE-2019-15914 (published 2019-12-20)\\
Description & ``An issue was discovered on Xiaomi DGNWG03LM, ZNCZ03LM, MCCGQ01LM, WSDCGQ01LM, RTCGQ01LM devices. Attackers can use the ZigBee trust center rejoin procedure to perform multiple denial of service attacks.''\\
CWE & CWE-20 (Improper Input Validation)\\
CVSS v3.1 & 7.5 (AV:N/AC:L/PR:N/UI:N/S:U/C:N/I:N/A:H)\\
CPE records & 
\texttt{cpe:2.3:o:mi:zncz03lm\_firmware:-:*:*:*:*:*:*:*} \newline
AND \texttt{cpe:2.3:h:mi:zncz03lm:-:*:*:*:*:*:*:*} \newline
\texttt{cpe:2.3:o:mi:mccgq01lm\_firmware:-:*:*:*:*:*:*:*} \newline
AND \texttt{cpe:2.3:h:mi:mccgq01lm:-:*:*:*:*:*:*:*} \newline
\texttt{cpe:2.3:o:mi:wsdcgq01lm\_firmware:-:*:*:*:*:*:*:*} \newline
AND \texttt{cpe:2.3:h:mi:wsdcgq01lm:-:*:*:*:*:*:*:*} \newline
\texttt{cpe:2.3:o:mi:rtcgq01lm\_firmware:-:*:*:*:*:*:*:*} \newline
AND \texttt{cpe:2.3:h:mi:rtcgq01lm:-:*:*:*:*:*:*:*} \\
CPE format & \texttt{cpe:2.3:\textit{part}:\textit{vendor}:\textit{product}:\textit{version}:\dots}\\
\bottomrule
\end{tabular}

\end{table}

\begin{figure*}
    \centering
    \includegraphics[width=0.97\linewidth]{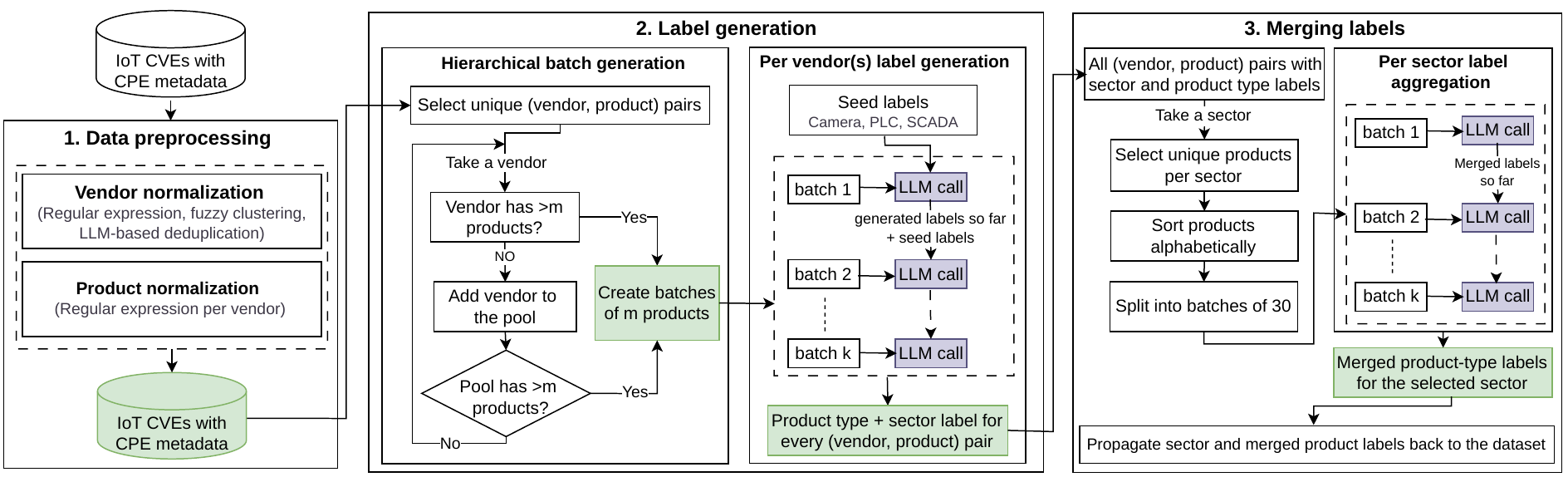}
    \caption{Methodology overview.}
    \label{fig:methodology}
\end{figure*}
\section{An overview of methodology}
\label{app:methodology}
Fig.~\ref{fig:methodology} summarizes the pipeline described in Section~\ref{sec:methodology}: (1) vendor and product normalization, (2) hierarchical batching and per-vendor LLM label generation, and (3) per-sector merging of the generated labels, which are then propagated back to the dataset.

\section{Prompts for Product Type and Sector Label Generation}
\label{app:prompts}
We provide the system and user prompts used by our LLM-based clustering pipeline for assigning product type and sector labels to each (vendor, product) pair.

\begin{promptbox}{System and user prompts for product type and sector label generation}
\label{box:sysprompt}

\textbf{\rmfamily System prompt:}

You are an expert in IoT and cybersecurity asset classification. Your task is to analyze CVE records and classify them.

Each CVE describes a vulnerability that affects one or more products. As an example, a vulnerability caused by MQTT might affect cameras, power meters and smart lights. Your task is to tag each of the affected products of a CVE by (1) the product type and (2) the sector they affect.

\textbf{Product type tagging guidelines:}
\begin{itemize}
    \item READ THE CVE DESCRIPTION FIRST, which often provides information about the affected products.
    \item USE VENDOR CONTEXT. The vendor's business is a strong prior. A product from Foscam is most likely a camera.
    \item Analyze product names and NAME THE PRODUCT TYPE by what the product physically IS or what function it performs, not by the technology inside it or the interface it exposes.
    \item Tag names must be vendor-agnostic and function-specific.
    \item Reuse an existing tag ONLY if it precisely describes the product's function. Do not reuse a generic tag because it loosely fits.
    \item Create a new tag if no existing tag precisely fits. New tags are encouraged when needed.
\end{itemize}

The following labels are too generic and are forbidden. If you find yourself wanting to use one of these, you must either pick a more specific tag from the existing list, create a new specific tag, or output 'UNCERTAIN':
\begin{itemize}
    \item 'IoT Device', 'IoT Gateway', 'IoT Platform', 'IoT Software', 'IoT Firmware'
    \item 'Consumer IoT Device', 'Industrial IoT Device'
    \item 'Software', 'Firmware', 'Middleware', 'Web Application', 'HTTP Server'
    \item 'Embedded Device', 'Hardware', 'Chip', 'Module'
    \item 'Protocol', 'Diagnostics', 'Field Device'
\end{itemize}

\textbf{Sector tagging guidelines:}
\begin{itemize}
    \item Choose sectors based on where the product is typically deployed/used.
    \item A product may belong to MULTIPLE sectors; include every sector that applies, but only if genuinely affected. Do not include loosely related sectors.
    \item Order sectors by relevance: list the PRIMARY sector first (most affected/deployed), followed by any additional sectors in decreasing order of relevance.
    \item Use ONLY sector names from the allowed list, spelled EXACTLY as given.
    \item Use 'Other' only when none of the listed sectors fit.
\end{itemize}

Return ONLY valid JSON.

\textbf{\rmfamily User prompt:}

CVE Information:\\
\{cve\_sections\_str\}

Affected Products: \{products\_str\}

Allowed sectors (use these names EXACTLY; assign one or more per product):\\
\{allowed\_sectors\_str\}

Existing product type tags:\\
\{clusters\_str\}

Do as follows: go through each affected product and follow the product type tagging guideline to tag the type of the product, and read the sector tagging guideline to determine the affected sector.

Return JSON with this exact structure:

\{
\hspace*{1em}"product\_mapping": \{\\
\hspace*{2em}"product\_name": \{"cluster": "cluster\_category", "sectors": ["sector1", "sector2"]\},\\
\hspace*{2em}"product\_name2": \{"cluster": "cluster\_category", "sectors": ["sector1"]\}\\
\hspace*{1em}\}
\}

Map EVERY product. Respond ONLY with raw JSON. No markdown, no explanations.
\end{promptbox}
\vspace{0.5em}

\begin{promptbox}{System and user prompts for merging labels}
\label{box:mergeprompt}
\textbf{\rmfamily System prompt:}
You are an expert in IoT functionality classification. You merge product type labels that describe the same kind of device or software, while keeping unrelated items separate. Be concise and consistent. Never use a popular label as a dumping ground for things that don't fit elsewhere.

\vspace{0.6em}
\textbf{\rmfamily User prompt:}

Here are product types from the `\{sector\}' sector.

PRODUCT TYPES TO MERGE:\\
\{pt\_str\}\\
\{committed\_str\}

RULES:
\begin{itemize}
    \item Merge entries that refer to essentially the same kind of product.
    \item Do NOT use a popular label as a catch-all. If an input does not clearly fit an existing cluster (e.g.\ a drone is not a camera, a music player is not a camera), give it its own minimal label.
    \item Use consistent canonical names (Title Case, no underscores).
    \item Every input product type must appear as a key in the mapping.
\end{itemize}

Respond with ONLY this JSON (no markdown, no code fences, no extra text):

\{\{\\
\hspace*{1em}``mapping'': \{\{\\
\hspace*{2em}``original\_product\_type'': ``merged\_category'',\\
\hspace*{2em}``another\_product\_type'': ``merged\_category''
\hspace*{1em}\}\}\\
\}\}
\end{promptbox}

\end{document}